\documentclass[utf8]{frontiersSCNS} 

\usepackage{url,lineno,microtype}

\usepackage[
  breaklinks = true,
   colorlinks = true,
  urlcolor   = blue,
  citecolor  = blue,
  linkcolor  = blue,
]{hyperref}

\newcommand{\aap}{    {\it Astron. Astrophys.}}

\newcommand{\apj}{    {\it Astrophys. J.}}
\newcommand{\apjl}{   {\it Astrophys. J. Lett.}}
\newcommand{\apss}{   {\it Astrophys. Space Sci.}}

\newcommand{\mnras}{  {\it Mon. Not. Roy. Astron. Soc.}}
\newcommand{\nat}{    {\it Nature}}

\newcommand{\pasj}{   {\it Pub. Astron. Soc. Japan}}

\newcommand{\solphys}{{\it Solar Phys.}}
 
\newcommand{\ssr}{    {\it Space Sci. Rev.}} 

\def\keyFont{\fontsize{8}{11}\helveticabold }
\def\firstAuthorLast{Brigitte Schmieder} 
\def\Authors{Brigitte Schmieder\,$^{1,2,3,*}$ }
\def\Address{$^{1}$ LESIA, Observatoire de Paris, Universit\'e PSL, CNRS, Sorbonne  Universit\'e,  Universit\'e de Paris, 5 place Jules Janssen, 92190 Meudon, France \\
$^{2}$ Centre for mathematical Plasma Astrophysics, Dept. of Mathematics, KU Leuven, 3001 Leuven, Belgium\\
${3}$  University of Glasgow, Glasgow, Scotland }
\def\corrAuthor{Corresponding Author}

\def\corrEmail{brigitte.schmieder@obspm.fr}

\begin{document}
\onecolumn
\firstpage{1}

\title[Solar jets: SDO and IRIS observations]{Solar jets: SDO and IRIS observations in the perspective of  new MHD simulations} 

\author[\firstauthor] {\Authors} 
\address{} 
\correspondance{} 

\extraAuth{}

\maketitle

\begin{abstract}
Solar jets are observed  as collimated plasma beams over a large range of temperatures and wavelengths. They have been observed in  
H$\alpha$ and optical lines for more than 50 years and called surges. The term "jet" comes from  X-ray observations after  the launch of the Yohkoh satellite  in 1991.
They are the means of transporting energy through the heliosphere and participate to the  corona heating  and the
acceleration of solar wind.
Several characteristics have been derived about their velocities, their rates of
occurrence, and their relationship with CMEs. However, the initiation mechanism of jets,
e.g. emerging flux, flux cancellation, or twist, is still debated.
In the last decade coordinated observations of the Interface Region Imaging Spectrograph (IRIS)  with the instruments on board  
the Solar Dynamic Observatory (SDO) allow to  
make a step forward for understanding the trigger of  jets and the relationship between hot jets and cool surges.
We observe at the same time the development of 2D and 3D MHD  numerical simulations to interpret the results. This paper summarizes recent studies of jets showing the loci of magnetic reconnection in null points or in bald patch regions forming a current sheet.  In the pre-jet phase a  twist is  frequently detected by the existence of  a  mini filament  close to the dome of emerging flux. 
The twist can also be  transferred  to the jet from a flux rope in the vicinity of the reconnection by slippage of the polarities.  Bidirectional flows are detected at the reconnection sites.  We show the role of magnetic currents detected in the footprints of  flux rope and  quasi-separatrix layers  for  initiating the jets.
 We select a few studies  and show that with the same observations,  different  interpretations  are possible based on different  approaches {\it e.g.} non linear force free field extrapolation or  3D MHD simulation.

\tiny
 \keyFont{ \section{Keywords:} solar jet, solar surge, solar flare, magnetic reconnection, EUV spectroscopy} 
\end{abstract}

\section{Historical studies}
Solar jets are  transient phenomena considered as  being  means of energy and mass transport in the solar atmosphere.
These are observed in multi temperatures {and wavelengths} from H$\alpha$ (for more than 50 years) to X-rays after  the launch of the Yohkoh satellite in  August 1991. Their kinematic  characteristics (velocity, acceleration, recurrence) have been derived using different space borne satellites and ground based observatories (see  recent reviews of  \citet{Innes2016,Raouafi2016,Hinode2019,Shen2021,DePontieu2021,Schmieder2022}).

Before describing the today's state-of-art of jets, let us look  at how surge and jet topic develops through  historical  papers leading to our present knowledge.
The development of instruments with higher and higher  spatial and temporal resolution  certainly helps us to make a step forward in our knowledge. So that  the cartoons proposed in the late 90's become Magneto-Hydro-Dynamic  (MHD) numerical simulations in 3-dimensions (3D).

\subsection{Spectroscopic analysis}
 Mass ejections such as sprays, eruptive prominences, and surges have been often observed in H$\alpha$ and other visible lines  \citep{Roy1973, Tandberg1974}. Most of the  results deduced from spectroheliograms or filtergrams concern the projected  trajectories of the material on the disk and the determination of velocity and acceleration.
 
 In the 1980's  new  surge observations were obtained using  spectroscopy from  space and ground instruments.   This is how the first Dopplergrams of jets were obtained     with the   Ultra-Violet Spectro-Polarimeter (UVSP) instrument on board the Solar Maximum mission in 1980  (SMM- \citet{Woodgat1980}) and consequently  new observational results    appear at that time.
 Coordinated  UVSP observations with the Multi Subtractive Double Pass (MSDP) spectrograph operating on the solar tower of Meudon  \citep{Mein1977}.
  allowed to obtain  full Dopplermaps (1 min x 8 min) in  H$\alpha$  and in  1548 \AA\ C IV  lines with the UVSP with a cadence of 30 s and a spatial resolution of 3 arcsec in  C IV and  1 arcsec  in H$\alpha$  respectively. Surges in H$\alpha$  appear as dark/absorbing  structures while they are bright  structures in emission in C IV. According to the low  spatial  resolution of the instruments both structures occupied the same area 
  and at the base of the surge a bright point was observed in CIV and H$\alpha$ \citep{Schmieder1982,Schmieder1983}.  In these  former studies of a surge   occurring on October 2, 1980,  its lifetime was around 20 minutes  with upflows followed by downflows,  with a  radial velocity reaching  60 km/s  in both lines H$\alpha$ and  C IV. Upflows and downflows were registered successively and not simultaneously like in rotating jets. However the wide line profiles in  C IV and H$\alpha$ could indicate  that along the line of sight  opposite flows exist  in unresolved structures. Nevertheless the large widths of  C IV line profiles  were
    explained by  a high microturbulence of 120 km s$^{-1}$. { In this context the micro-turbulence is the non-thermal microscopic component  of the gas velocity in the formation zone of spectral lines.   It is frequently used  to explain broadened line profiles in the stellar spectra.}
  H$\alpha$ dopplershifts (radial velocity) were computed with the  cloud model technique \citep{Mein1996,Gu1996},  and the horizontal speed by following the leading edge of  surges.  Measuring Dopplershifts  supports the idea that in surges   there are   mass motions and not propagating waves.  During the  initiation phase of  surges  flow acceleration was  established in  H$\alpha$ and C IV  which  permitted the authors to conclude that pressure gradients could be the driving force of the surge.  However the acceleration phases were different in both lines  which means that C IV  emission did  not come from  the transition region of  surges but come from independent structures.  C IV surge quantities (velocity, acceleration) varied on a very short time scale as if there were pinched zones in the magnetic tube.  
 {   In H$\alpha$  the displacement of the maximum velocity was observed  along the axis of the surge. In C IV   the velocity maxima are observed  at given distances along the surge    strong upward velocity  maximum  followed by low velocity with no propagation
during the evolution of the surge,
suggesting  the existence of kink waves.  This kind of behaviour for surges observed in transition region temperature  has not been  repeated  but has been observed  and modelled for  spicules \citep{He2009}. A Dopplershift signature  with blue and redshift  from one edge to the radially opposite edge for a given surge cross-section was interpreted as  torsional waves. However in these earlier observations   the authors  favoured the interpretation of  successive  upflows and downflows, implying no rotation. As noted in the chapter 3.1  torsional waves seem to be  more frequently observed.
  Torsional  waves   are used  now in MHD simulations  as drivers of jets \citep{Pariat2015}.}

  { Radio Type III  bursts were   also often associated with surges suggesting that surges  followed open magnetic field lines or very large loops}
  \citep{Chiuderi1986,Kundu1995}.  This idea has been confirmed by using NLFFF extrapolation showing  how non thermal types III  associated with jets  escape along open field lines at the  edge of close structures over  active regions  \citep{Mulay2019,Lu2019}.
\citet{Schmieder1983,Schmieder1984} reported on the recurrence of H$\alpha$ and C IV surges with a time delay between two jet ejections of 15 to 30 min.  They proposed that such recurrent ejections could be due to periodic energy storage and periodic reorganisation of magnetic field as envisaged to occur for flares, but at lower energy levels. 


\begin{figure*}[t!]
\centering
\includegraphics[width=0.75\textwidth]{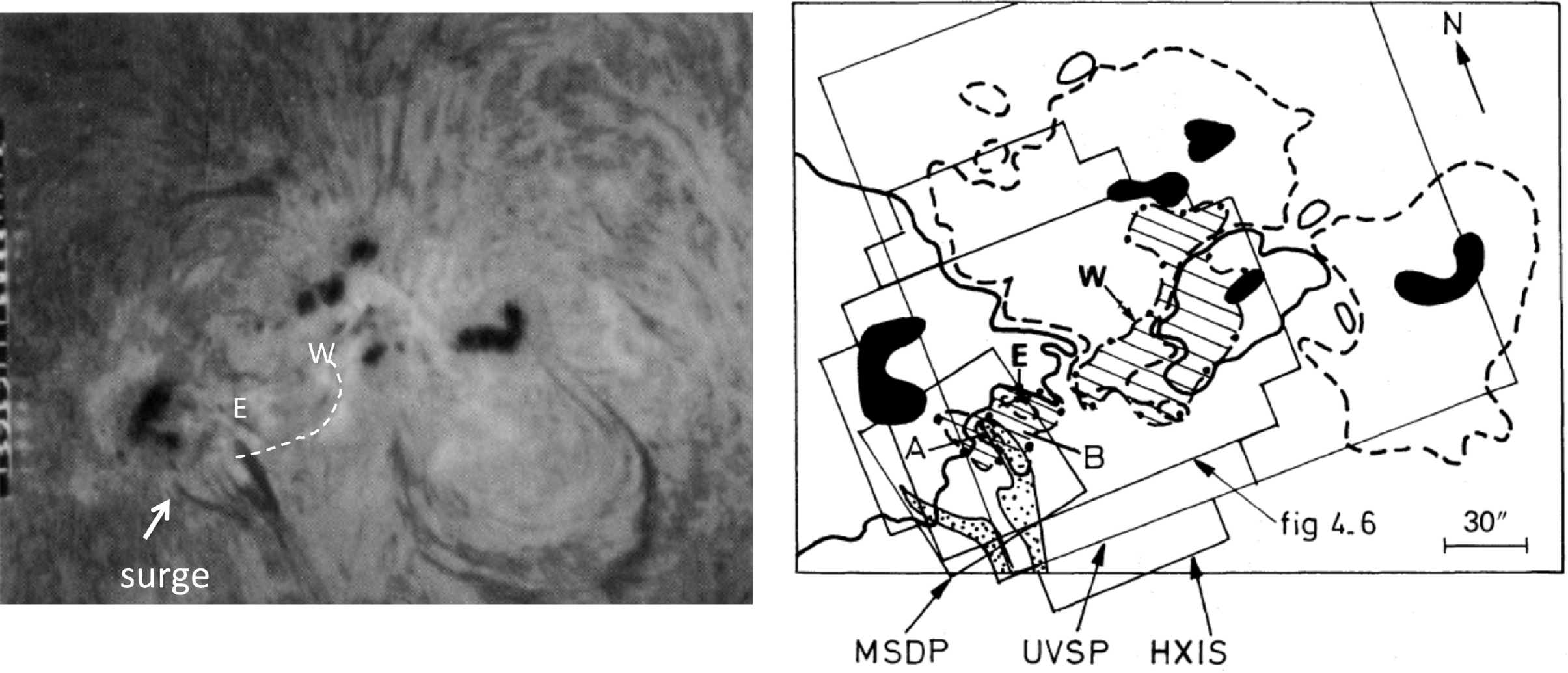}
\caption{Surge in H$\alpha$  observed on November 11 1980,  with the Solar  Optical Observations Network  (left panel) and all the signatures  of the event obtained  by the instruments (UVSP, HXIS) on board SMM (right panel).  The surge  in H$\alpha$ from the MSDP operating in the Meudon solar tower, and in O V  from the UVSP is indicated  by the  dotted area, {   the bright H$\alpha$ areas at its  footpoint  by the letters A and B}, the hatched area  represents the footpoints  (E, W) of a loop  observed  in soft X ray (3.5-5.5 eV) with HXIS  and Fe XXI   with the UVSP. The EW loop   is drawn with a dashed line in the H$\alpha$ image (adapted from \citet{Schmieder1988})
}
\label{Fig_schmieder}
\end{figure*}

\subsection{Energy budget in surges and X-ray loops}
{ The energy budget was determined  by  analysing    the  signatures of surges  and jets in multi-wavelengths and multi-temperatures obtained  by the instruments on board   the Solar Maximum Mission   (SMM)  launched in  1980. Coordinated  campaigns   with ground based  instruments  allowed   simultaneous observations } in H$\alpha$  with the MSDP operating on the solar tower in Meudon, in O V and Fe XXI with the UVSP/SMM and in soft X rays 
with the HXIS/SMM \citep{Schmieder1988,Schmieder1993,Schmieder1996_Sim}.
The cool   (H$\alpha$)   and  warm (O V) surge plasma show velocities of  the order of 120 km/s
in a  comb-shaped surge observed on November 11 1980 at the edge of a sunspot \citep{Schmieder1988}.
 The surge  intensity  (H$\alpha$ and OV) was well correlated with  the emission of an  associated loop   observed in the HXIS channel (3.5-5.5 keV) and in the FeXX1  line (UVSP),  one footpoint of the loop being close to the footpoint of the jet (Figure \ref{Fig_schmieder}).  This suggests that the surge could be due to the reconnection between  the closed loop and open field. 
 The  association of X-ray and UV emission with H$\alpha$ surges allowed  the authors to estimate the  energy budget between kinetic, potential and radiative energy.
 The potential and kinetic energy was  both of the order of 2.5 -5 10$^{28}$ ergs, two orders larger than the radiative loss in the X-ray loop. They concluded that the magnetic energy liberated at the base of the surge  was mainly transferred to kinetic energy and only a small part was released in thermal  and nonthermal energy.  In a successive paper analysing different multi wavelength data \citep{Schmieder1996_Sim}, the authors concluded that the magnetic reconnection should occur in the corona. The energy is transported by energetic particles along the loops. The energetic particles are losing energy in the chromosphere as surges in open field lines and as bright loops in close loops like for mini-flares.
 The X-ray  spikes  appear earlier than H$\alpha$ and UV surges.  The maximum upward velocity happens ten minutes after the onset of the surge. The response of the chromosphere depends on the magnetic topology.
 
 The  partition of magnetic energy released as kinetic or thermal energy  during reconnection between  close loops and open structures is  still not clear. A statistical analysis of the relationship between miniflare and jets show no positive answer and exhibits broad  distributions of the delay  between these two events and  of their amplitudes \citep{Musset2020}. This confirms the result of this historical study that   flares and  associated jets belong to a global system where the energy  is released during reconnection and the partition depends on the magnetic configuration of the system (close and/or open structures). { In \citet{Musset2020} the delay between the  non-thermal X-ray peak 
emission and the peak intensity  of the jet    is negligible, which   implies  that the 
jet could be  produced by magnetic reconnection. However this does not  give  any information on the magnetic configuration.}
 Pressure pulses can also be created by magnetic reconnection  which may release impulsively  energy  and heats the plasma in closed and open flux tubes. 
 { In the standard model  of magnetic reconnection for  eruption  the chromosphere plasma is heated and evaporates}.
 In closed flux tubes the density increases,  and  strong pressure and temperature gradients  produce upward motions \citep{Shimojo2001}. Indeed pressure-driven up-flows are slow; they correspond to  trans-sonic  flows. This model is may be still valid for surges but is not applicable to fast X-ray jets.  This  concept of pressure gradient for initiating surges  has been discussed in the context of  evaporation model and criticised because it  requires  heating the chromosphere to transition zone temperatures and then cooling. However the cooling time is very short  at these temperatures with no delay \citep{Schmieder1994b}.
 
\begin{figure*}[ht!]
\centering
\includegraphics[width=0.8\textwidth]{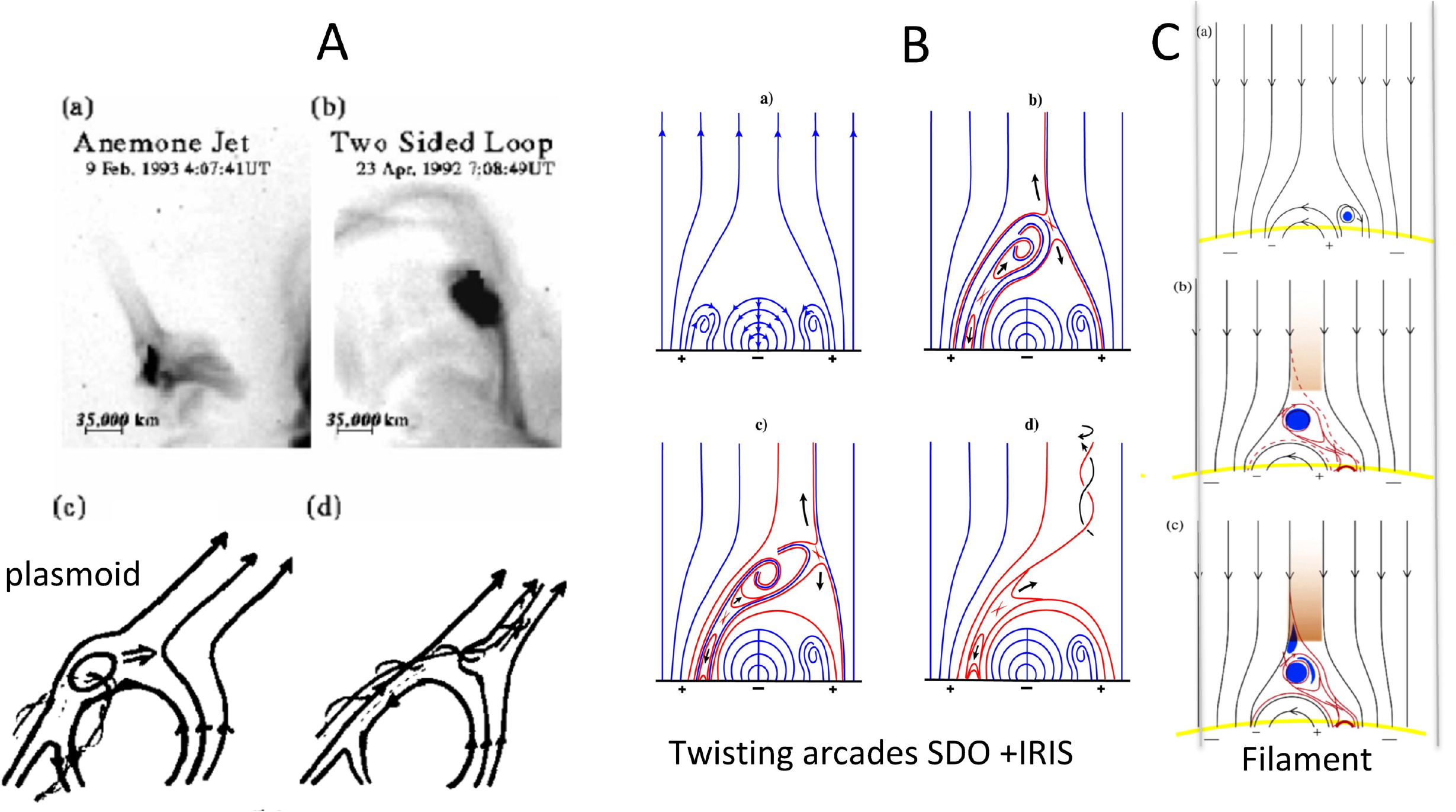}
\caption{Cartoons representing the initiation of jets. Left column (A): (panel a) two kinds of X-ray  jets observed with Yohkoh: anemone jet and two side loop jet, panel (b) unified  CSHSP model  of flares and flux emerging model: the plasmoid-induced reconnection model proposed by  \citet{Yokoyama1995,Shibata1998} (adapted  from \citet{Shibata1999}). The plasmoid is a magnetic island or a twisted flux rope. Middle column (B): a possible formation mechanism of penumbral jets from the eruption of a magnetic arcade (the core of which becomes a twisted flux rope as it erupts) inserting shear and twist in the jet spire. In this sketch only the "cross-cut/section" across the central axis of the tail of the penumbral filament is represented so that the field in the filament head is behind the viewer (in the line of sight) (adapted from \citet{Tiwari2018}).
 Right column (C): blowout jet initiated by a mini filament, in panel (b) two red crosses indicate 2 sites of reconnection: the low cross corresponds to tether cutting reconnection and a bright point:loop forms below, the top cross corresponds  to break-out (adapted from \citet{Sterling2015,Sterling2020}).
}
\label{Fig_cartoon}
\end{figure*}
 \subsection{X-ray jets}
 The  Yohkoh satellite launched on August 30 1991 with on board the SXT instrument \citep{Tsuneta1991}  helps  us to make definitively progress on X-ray jets and to  decide if jets was initiated by  pressure  pulse \citep{Sterling1994} or by magnetic reconnection
  \citep{Shibata1992,Shibata1994,Shibata1996}. The former authors  defined X-ray jets as transitory X-ray enhancements with  collimated motions. All the jets are associated to microflares. Their length is 1000 to 4 x 10$^5$ km. Their apparent speed is 10-1000 km/s, the temperature 4-6 MK. The morphology of X-ray jets show converging shape (lambda-shaped) suggesting a null point near the footpoint of the jet.  Parasitic polarities are often observed in the footpoint favouring magnetic reconnection and  this fact gives evidence of a null point. Surges could accompany X-ray jets  \citep{Canfield1996,Schmieder1996_Shi}. In the former paper  the association with the jet was an X-ray loop and not a fine  X-ray jet similarl to the previous observations with HXIS/SMM
  \citep{Schmieder1996_Sim}. The hot footpoint is not always exactly at one end of the jet and could be represented by a loop.

  A unified  reconnection model valid for flares and jets called plasmoid-induced-reconnection  was proposed by Shibata  \citep{Yokoyama1995,Shibata1999} (Figure \ref{Fig_cartoon} left column A, panels c and d).  The standard  CSHKP model  and the emerging flux model were compatible with this plasmoid-induced-reconnection where plasmoids were compared to flux rope ejected during flare.  { Therefore they proposed that} reconnection occurred between the plasmoids and the ambient field, and hot loops formed below like post-flare loops  as in the standard  flare model.  Simulations in 2 and 2.5  dimensions develop possible models based on the  conceptual idea that 
  magnetic reconnection   may accelerate the plasma in two   ways \citep{Shibata1985,Shibata1986,Shibata1996}.  With  the  tension-driven  model,  plasma is  accelerated  to  Alfv\'enic  velocities  in  the  vicinity  of  the  reconnection  site  as a sling-shot mechanism \citep{Yokoyama1996,Moreno2008}.  The second way is characteristic of the initial magnetic configuration:  the  untwisting model in which  the closed magnetic structures should possess initially shear or twist  \citep{Schmieder1995,Canfield1996,Jibben2004}. These two concepts are important in the acceleration process of  jets  to reach Alfv\'enic  velocities or sub-Alfv\'enic  velocities.
{ Recent papers appear to favour the shear  inside the embedded structure  from which jets will be initiated \citep{Kumar2019}.}  { In 3D  simulation it  is clear that  plasmoids are created during reconnection and are ejected along current sheets   as it is shown in  \citet{Kumar2019}. }  { In 3D  the  curvature of magnetic field lines can be in both directions  therefore the distinction of these two mechanisms is not clear.  In  3D both mechanisms contribute simultaneously to driving plasma from reconnection sites. Some amount of shear or twist in one component of the reconnecting flux systems is needed to provide enough free energy for the eruption/jet; otherwise, as many models have shown, only weak jets result.}

 \subsection{New instruments (2000-2010)} 
  
 Later high spatial resolution instruments  were developed  and brought new imaging and spectral observations.
{  Solar jets were observed in different regions of the sun: network, coronal hole, active region, in the chromosphere,  and in the corona.   Theory  and interpretation  were rapidly developing.  We  just  list the new generation instruments with  some relevant papers, such as }
the Swedish solar telescope  (SST) \citep{Nobrega2017},  { the NST/GST  at the Big Bear Observatory \citep{Kumar2015}}, the TRACE mission with its UV instrument  \citep{Alexander1999}, Hinode  with its SOT polarimeter,  its spectrograph EIS \citep{Muglach2021} and XRT \citep{Madjarska2011}  showing   multi-wavelength jets with  different spatial, physical and temporal properties in  coronal holes  and  quiet sun. 
{  \citet{Hinode2019}  (section 7) summarises significant progress  with the insightful observations   using  the advanced instruments (e.g., EIS, SOT and XRT)  \citep{Cirtain2007,Savcheva2007,He2009}.}

 { For example we may note the  detection of the excitation and launch of kink waves due to magnetic reconnection.
  An other example of the merit contributed by EIS and XRT for  coronal jet study  was  the observation of  a mini-CME  \citep{He2010}.   
Time-varying Dopplergrams of a mini-CME event were successfully captured and recorded by EIS when it was repetitively rastering the same solar region with a repetition period of about 6 min. The initial eruption speed was estimated to be as low as 30 km/s from the Dopplergram. The associated X-ray emission of the mini-CME started from a sudden brightening at one footpoint of the closed loop and then a rapid propagation of the brightening along the erupting closed loop. This loop could represent the onset of a mini-CME.}
 
\begin{figure*}[t!]
\centering
\includegraphics[width=0.8\textwidth]{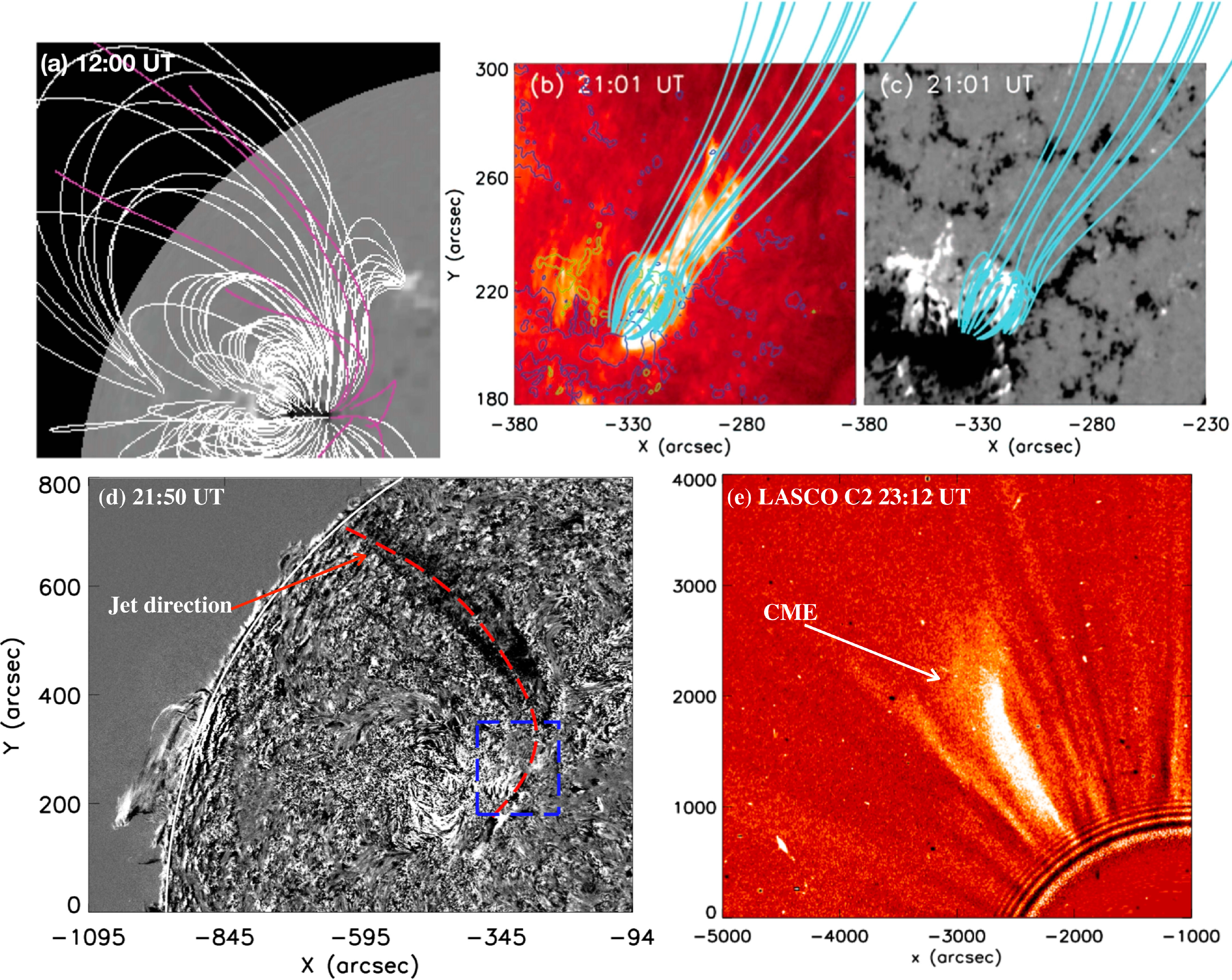}
\caption{Jet initiating a narrow CME:  PFSS extrapolation of the large field of view (FOV) is shown in panel (a). The white and pink lines are the closed and open magnetic field lines at the jet
location. The open field lines mirror the jet propagation from its source to the solar corona, which is indicated by a curve  (dashed red line   in panel d). In panels (b) and (c)  are presented the  AIA 304 Å jet observation  and the   HMI magnetogram, respectively, showing the jet source. 
The cyan lines  are the
magnetic field lines from the source region, which shows a closed structure at the jet base and open lines afterward. 
Panel (e) shows   the narrow CME observed by LASCO C2 (adapted from \citet{Joshi2020CME}.
}
\label{Fig_CME}
\end{figure*}

 Later on   many examples  showed that  collimated jets can produce  coronal mass ejections  observed with coronagraphs e.g. SOHO/LASCO  (Figure \ref{Fig_CME})
 \citep{Sterling2018,Joshi2020CME,Panesar2016b, Kumar2021} and acceleration of particles \citep{Mulay2018,Joshi2021b}.

  In the following years the  more important  achievement was due to 
  the development of theory to reply to the following questions: what is the driver of jet? what is the relationship between jet and surge?
 What are the physical conditions of jet  and surge?
  We  focus the next sections on recent observations using SDO and IRIS.   
  This development has its seeds in the   observations of  many jets  after the launch of SDO  in 2010 and more recently the launch of  the Interface Region Imaging Spectrograph (IRIS -\citet{Pontieu2014})  in 2013. 
 
 In  Section 2 are presented the results of SDO observations which inspire the development of   3D MHD simulations based on flux emergence as trigger of jet.
 In Section 3 are presented twisted jets observed by SDO and IRIS, and  their interpretation using 3D simulations based on the existence of the transfer of   twist.

\section{SDO and the onset of jets}
 In 2010 SDO was launched with onboard two important instruments
 the Atmospheric Imaging Assembly \citep[AIA:][]{Lemen2012} providing 
 extreme ultraviolet (EUV) and ultraviolet (UV) data, and  the Helioseismic and Magnetic Imager \citep[HMI:][]{Scherrer2012} providing magnetograms both  with a high spatial resolution (0.6 and 0.5 arc sec respectively) and high temporal cadence (12 sec { and 45 sec respectively}).
 A very impressive number of  publications concerning  jets observed with AIA with more precise results of  their characteristics and their triggers appeared 
 \citep{Raouafi2016,Shen2021}.  { In the review of \citet{Hinode2019} there is an interesting discussion of the  jets observed with  the Hinode instruments but also  a deep discussion on  the origin of the jets  observed with SDO/AIA.  }
 \subsection{Morphology of jets}

With Yohkoh  observations
  coronal  jets were  classified   in two types:  straight  anemone  jets and  two-sided-loop  jets  (Figure \ref{Fig_cartoon} left column A, panels a, and b) \citep{Shibata1994}. Anemone jets consist of a collimated jet and a dome-like base corresponding to  magnetic flux emergence. The two-sided-loop jets exhibit diverging flows from their  excitation center. This new kind of jet was also observed with Hinode /XRT and EIS instruments, which confirmed
the opposite direction flows  via Doplershifts  \citep{Sterling2019}.

With SDO/AIA   new kind of jet was discovered called blowout jets \citep{Moore2010}. Compared to standard jets,  they exhibit different characteristics:  an additional  bright point inside the dome, a blowout eruption of the base arch that could host a twisting mini-filament, and an extra jet-spire strand  close to the  outside bright point.  The probability  of a jet to be  a blowout jet  was found to reach  50 \%  \citep{Moore2013,Chandra2017}. They look like  break-out eruptions initiating CMEs \citep{Joshi2020,Kumar2021}.

Several observations show that mini-filament eruptions are closely associated with coronal jets and could be the triggers of blowout jets as in large scale filament eruptions before flares  (Figure  \ref{Fig_cartoon} right  column C) \citep{Sterling2015,Shen2012,Shen2017}. New cartoons have been proposed for the   magnetic configuration for jets  with a filament close to an emerging flux  (Figure  \ref{Fig_cartoon} right column C) \citep{Sterling2018,Sterling2020} and  
for jets in penumbra with multi-arcades configuration (Figure  \ref{Fig_cartoon} middle  column B) \citep{Tiwari2018}. 
In these observations  the origin of the jets is identified in magnetic flux cancellation rather than  in flux emergence \citep{Moore2010,Panesar2020}. 
{  \citet{Kumar2019} analysed 27 jets in equatorial coronal holes  using SDO/AIA  observations and found a high proportion of  jets involving  a  filament channel  eruption and  free  energy resulting from  shear motions or pre-existing twist, with no evidence of flux cancellation;. }

 \subsection{Flux cancellation}
{  Several studies show clearly that  tiny coronal jets in coronal holes and quiet sun   are due to flux cancellation \citep{Panesar2016a,Panesar2017,Panesar2018a,Savcheva2007,McGlasson2019}.
{   In coronal holes converging flows toward the boundary of super-granule with mixed polarities  leading to cancelling flux was found to be a  favourable  solution  to explain onset of jets \citep{Young2014,Muglach2021}.} 
\citet{McGlasson2019} made some statistics on 60  such coronal jets  observed with the AIA 171 filter and found that nearly all  are associated  with dark absorbing features between two opposite polarities  considered as proxies of  mini-filaments.   Jets  were    formed by   flux cancellation. They { said}  that it is the result of lower reconnected loop submergence into the photosphere.  
{ Two {bright} points were observed, one internal 
 brightening and one external   
 brightening 
  with extended magnetic field lines along which the jet was running. By analogy with the cartoon in  Figure 2 panel C they discussed that these two brightenings may correspond to the two reconnection points of this scheme.} These  coronal  jets lasted around 10 to 12 minutes. Their bases are  small between 8000 km and 17000 km in the case of the 10  coronal jets  analysed by \citet{Panesar2016a}.
  Similar pattern of jets  at the edge of network were also observed  simultaneously in AIA 171 \AA\  and in the slit-jaw images of IRIS  1400 filter containing Si IV lines \citep{Panesar2018b}.  They were triggered by the reconnection   due to cancelling }
  
 However the identification of  mini-filaments is sometimes questionable because  an arch filament system  (AFS) over emerging flux looks like  mini-filament (cool material). However their formation and magnetic configuration are completely different. mini-filament is along the inversion line (PIL)  between positive and negative polarities, AFS are perpendicular to the PIL and  are unsheared; they are not twisting filaments, so their free  energy iin that case. The mini-filament should be detected along the PIL between the dome of the emerging flux and the ambient field as shows the cartoon as shown in Figure  \ref{Fig_cartoon} (right column C). We show in subsection 2.3 an example of possible misinterpretation.

 {  The detection of cancellation of flux depends crucially on the spatial resolution of the  magnetograms and certainly HMI is not enough sensitive to detect small dispersed magnetic field and validates really cancellation of flux.  Reconnection between magnetic field lines is due to the motions of their footpoints  induced by the  convection. Therefore reconnection may occur in the whole corona depending on the magnetic topology of the region.
{  \citet{Kumar2018} showed  jet onsets resulting from explosive breakout reconnection  between the flux rope  inside the closed structure and the external open field, in  a
classic fan-spine magnetic topology, characterised by a slowly rising EUV-bright sigmoid
and mini-filament, dimmings at both ends of the sigmoid, weak
quasi-periodic outflows at the null,  and multiple plasmoid formation
in the flare current sheet beneath a rapidly rising flux rope.
There was no evidence of flux emergence or cancellation up to 16 hr before
the impulsive event. For this case, the observed features closely
matched the predictions of  breakout-jet models \citep{Wyper2018,Wyper2019}. }}
 
\begin{figure*}[ht!]
\centering
\includegraphics[width=0.8\textwidth]{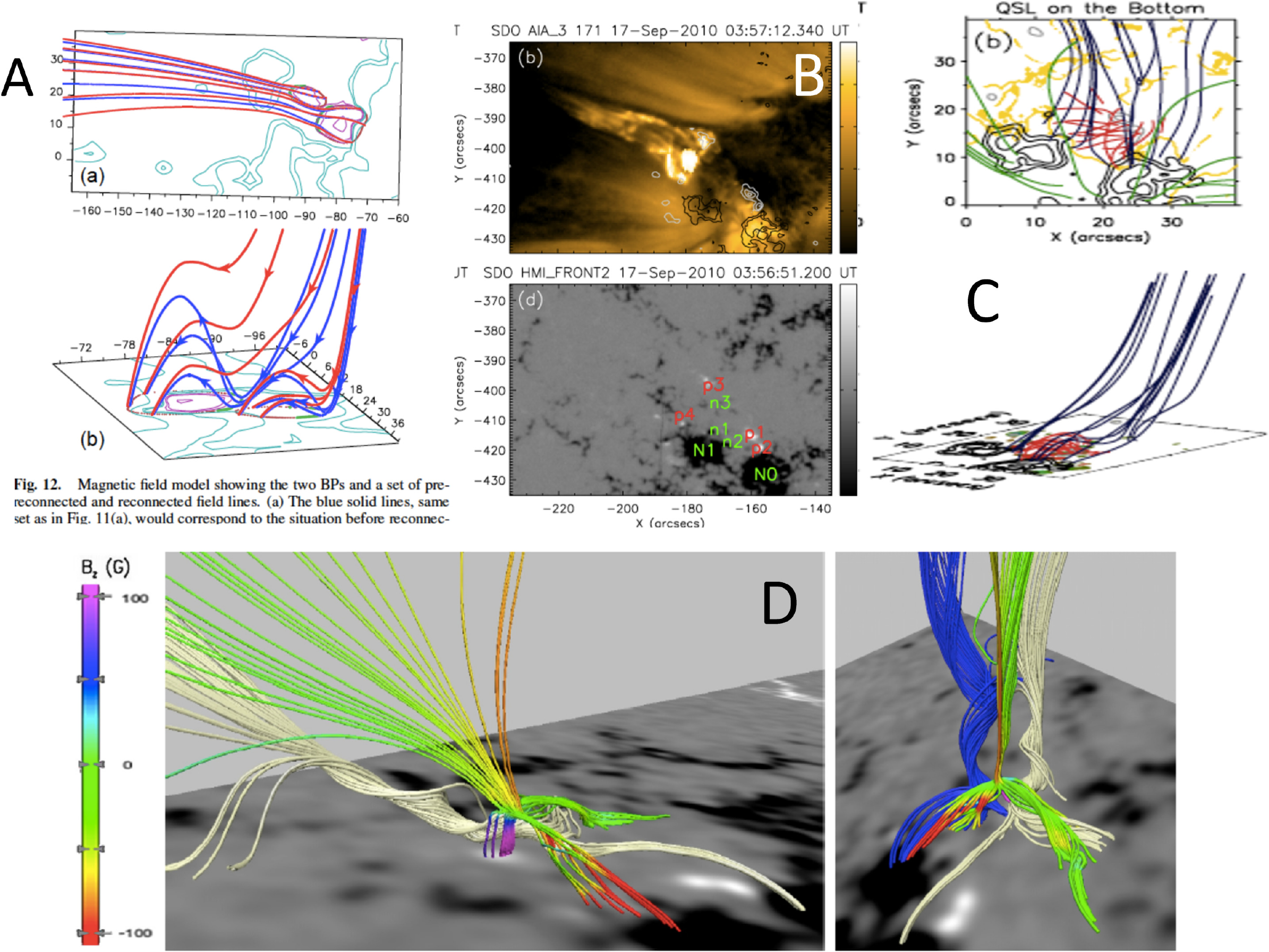}
\caption{Magnetic reconnection initiating solar jets  in bald patch  (BP) and in null point. Top left panels (A) reconnection  by bald patches, viewed from above and side view (from \citet{Chandra2017}) ; Middle panels (B): jet observed by AIA 171 \AA\ filter and the corresponding HMI magnetic map ;  Right panels (C) : NLFF extrapolated magnetic field  lines with bald patch region and electric currents  (from \citet{Guo2013}). Bottom panels:  NLFF  extrapolated magnetic field showing bald patch and null point (from \citet{Schmieder2013}).
}
\label{Fig_chandra}
\end{figure*}

\subsection{Flux emergence, null point and bald patch}


Concerning  magnetic flux emergence it is currently accepted that a  null point or separator is formed   between the emergence and the surrounding  magnetic field. During      reconnection at  null point,    energy can be  released in the form of a  flare, eruption, or jet
\citep{Filippov1999}. In the corona the magnetic field is free and frozen into the plasma almost everywhere. Only  at  null points and in current sheet s (frequently  present in separatrices)  can the energy release occur.
A  magnetic  field configuration  with separatrices can favor the occurrence of jets. For example it was shown that   magnetic field lines over  emerging flux of  a bipole  close to a sunspot   could reconnect to the ambient open magnetic field lines via bald patch (BP) regions \citep{Guo2013,Chandra2017}. BPs are regions where the magnetic field lines are tangent to the solar surface
{ (see Figure \ref{Fig_chandra}  top left and right panels  where  blue/red lines  are   for the pre-reconnected/reconnected  magnetic field lines respectively). 
Blue magnetic field lines bend towards the photosphere  as they are attracted by opposite polarities. { They are pinched together and reconnect  to form the red magnetic field lines  which are no more tangent to the solar surface.} In a BP  the magnetic field  appears to be   going from negative  to positive polarity contrary to loops where magnetic field is going from positive to negative. With shear motions the two branches of the BP insert a  thin layer  like in  separatrix and electric currents are stored until  reconnection occurs and  releases  the energy.}

In the \citet{Guo2013} study, a  jet   observed on September 17 2010 by AIA and HMI expanded in 10 minutes to 100 Mm in length with a speed of 200 km/s and a large base.
 BPs have been found in a  non linear force free extrapolation of photospheric magnetograms. { It was proposed that  magnetic reconnection could occur  at the BP separatrices.
  During the reconnection cool plasma
   could be ejected  along  open field lines driving jets.} This kind of evolving magnetic structure called
separatrices or quasi-separatrix layers  (QSL) are known to be  the location of drastic changes of connectivity and narrow current layers are created along them \citep{Demoulin1996}. In the case of \citet{Guo2013} QSL footprints  with electric current   were detected   around the emerging bipole base of the jet 
close to  the main polarity  (see Figure \ref{Fig_chandra} right panels).  { The recurrence of the jets was co-temporally related to the accumulation of electric currents}.
{   In this study  it was shown that these jets could be explained by flux emergence \citep{Shibata1998} and also by the converging flux model \citep{Priest1994} since the newly emerged magnetic flux is consistent with the former model and the bald patch configuration is consistent with the latter one. But both models are two dimensional with magnetic reconnection in separatrices  (as implied by their dimensionality), while the magnetic connectivity is not necessarily discontinuous in the three dimensional space. The above models can be generalised as a three-dimensional configuration with a magnetic null point (e.g., \citet{Moreno2008,Torok2009,Pariat2009,Wyper2018}.}

Revisiting  these data,  and analysing carefully the
non-linear force-free extrapolation, BP separatrices, low-altitude flux ropes,  and  even a null point were identified at the base of the jet (see Figure \ref{Fig_chandra} bottom panels) \citep{Schmieder2013}. 
Therefore we conclude that it is difficult to identify  clearly in 3D magnetic  field extrapolations  the  exact regions where energy is evacuated. Commonly many low altitude null points and separatrices exist and are the  possible sites of triggering jets and eruptions.  
 Nevertheless while  standard  or Eiffel-tower-shaped jets  appear to be caused by reconnection in current sheets containing null points,  reconnection in regions containing bald patches, such as 
   the jet of \citet{Guo2013,Joshi2020FR} seems to be of prior importance for triggering the jet. 

\begin{figure*}[t!]
\centering
\includegraphics[width=0.8\textwidth]{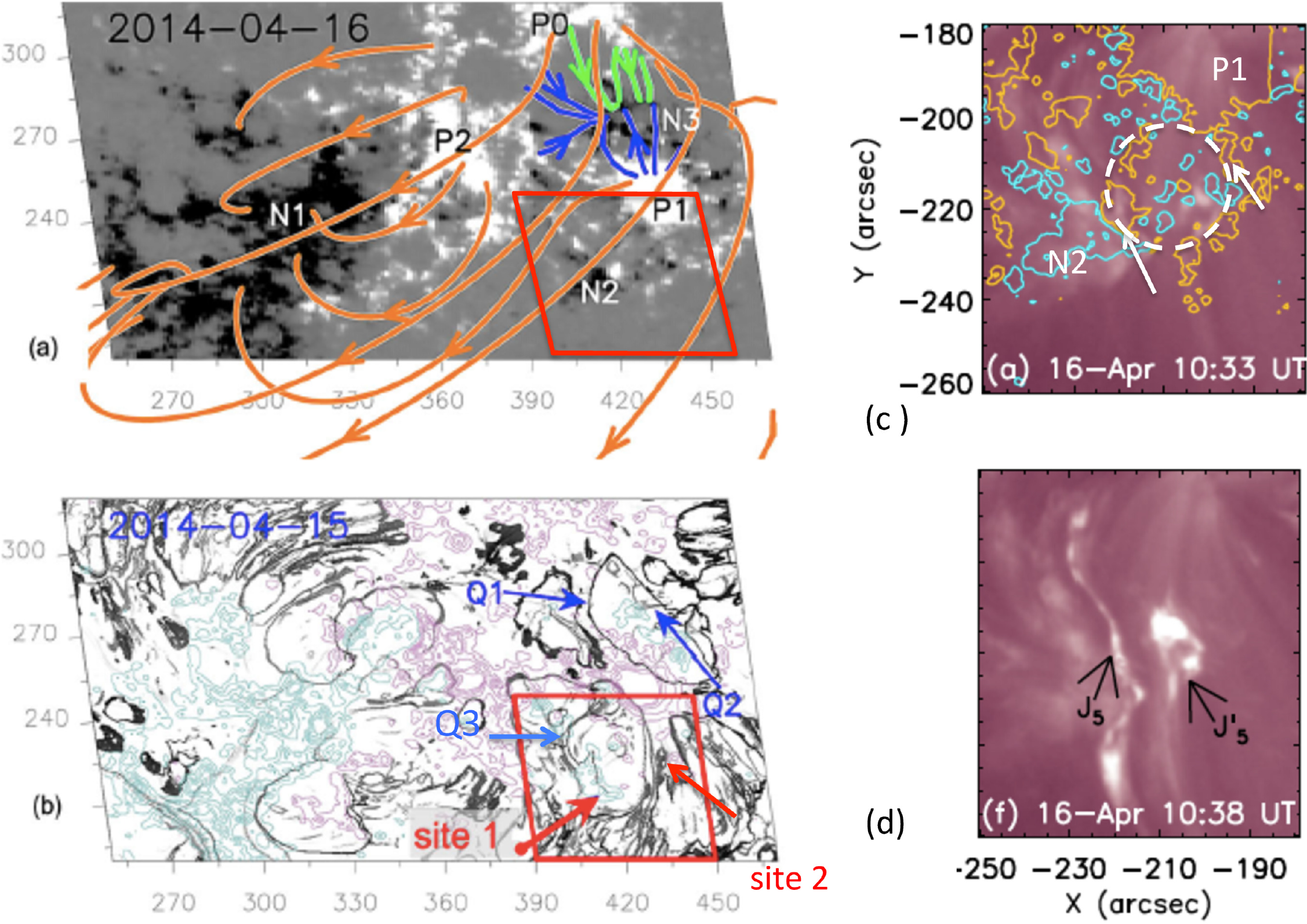}
\caption{Active Region  NOAA  12035 observed on April, 15-16 2014, (a) HMI magnetogram overlaid by extrapolated magnetic field lines, (b)  Quasi separatrix layers (QSL)  over the magnetic field contours, the red box indicates the  2 sites of the reconnection of the jets
(red arrows). (c)  Zoom on the magnetic region  formed by mixed emerging polarities (inside the  white circle) between polarities  P1 and N2 , (d) the two parallel jets with origin in the 2 sites (adapted from \citet{Joshi2017}).
}
\label{Fig_joshi2017}
\end{figure*}

{   The question is,  what is the role of flux emergence compared to flux cancellation role.} 
In view  of this divergence of opinions about the main trigger of the jets, it is difficult to  determine which mechanism dominates \citep{Schmieder2014}.
{  In fact the main trigger is the convection  which moves the magnetic field lines in the solar atmosphere and  leads to  reconnection. Convection is responsible for emergence, cancellation, shear, twist 
and finally generates free energy. Motions of footpoints of magnetic field lines allow to accumulate electric current   in the special loci where magnetic field can change easily of connectivity, in the QSLs. Before  jet onsets, frequently bright points are observed in the footprints of the QSLs. When enough energy is stored as  it was shown with the measurement of electric currents \citep{Guo2013},  the release of energy produces kinetic phenomena such as  jets or eruptions.   In the breakout model,  reconnection  between the flux rope below the breakout current sheet can produce very energetic  events. }

\subsection{Case study of emerging flux on the disk}

This  Section   explains how, using the same data of jets observed with AIA and HMI,  two different groups concluded differently on the trigger of the jets.
One group explains the jets  as slipping reconnection around  flux emergence \citep{Joshi2017}. 
The second group  explains the jet as a blowout jet driven by  the eruption of a mini-filament \citep{Shen2017}.
In fact eleven recurring solar jets  originated from two
different sites (site 1 and site 2) close to each other (about 11 Mm)  in the  NOAA active region
(AR) 12035 during 15 – 16 April 2014  \citep{Joshi2017}. The analysis of the   active-region  magnetic configuration showed that  a strong bipole  (P2-N2) was emerging  on April 15 2014  inside a remnant active region (P1-N1) (Figure \ref{Fig_joshi2017} panel a). In the neighbourhood of P1  flux  emergence  continuously occurred  between P1 and N2 and a circle-shaped  quasi separatrix layer (QSL),  was  detected around these new  emerging polarities  (Figure \ref{Fig_joshi2017} panels b and c).  On 16 April, both sites are located in  QSLs,  Flux emergence and cancellation mechanisms   triggered  the eleven jets in site 1 and/or site 2.
The jets of both sites had parallel trajectories and moved to the south with a speed between
100 and 360 km s$^{-1}$. The jets of site 2 occurring during the second day had a tendency to
move toward the jets of site 1 and merge with them. It was  conjectured that the slippage of the
jets could be explained by the complex topology of the region, which included a few low altitude
null points and many quasi-separatrix layers (QSLs), which could intersect with one
another.

Only one  of these jets  at 07:40 UT has been analysed by \citet{Shen2017} using  SDO and the New Vacuum Solar Telescope (NVST - \citet{Liu2014})  in  Fuxian lake in China observations.  Their interpretation of the trigger of this jet is different  from the series of jet described above.  Effectively  with the high resolution of the NVST they could detect  in H$\alpha$  a dark absorbing feature  perpendicular to the jet direction. They explained the event as a blowout jet, like a mini eruption driven by the fibril that they call mini-filament.  The jet is mainly observed in hot plasma  (emission in the AIA 171) with no cool jet visible in absorption as the other series. It would mean that the active and complex emerging flux close to the sunspot triggers the  recurrent jets in different ways. The site of reconnection is already slipping along the QSL, so   null point reconnection could be more important for this blowout  jet that is concerned.  On the other hand 
the dark structure  could also be an arch filament system and not a real mini-filament with no twist.  The interpretation of observations is very complex and we need more and more high resolution instruments.

\begin{figure*}[t!]
\centering
\includegraphics[width=\textwidth]{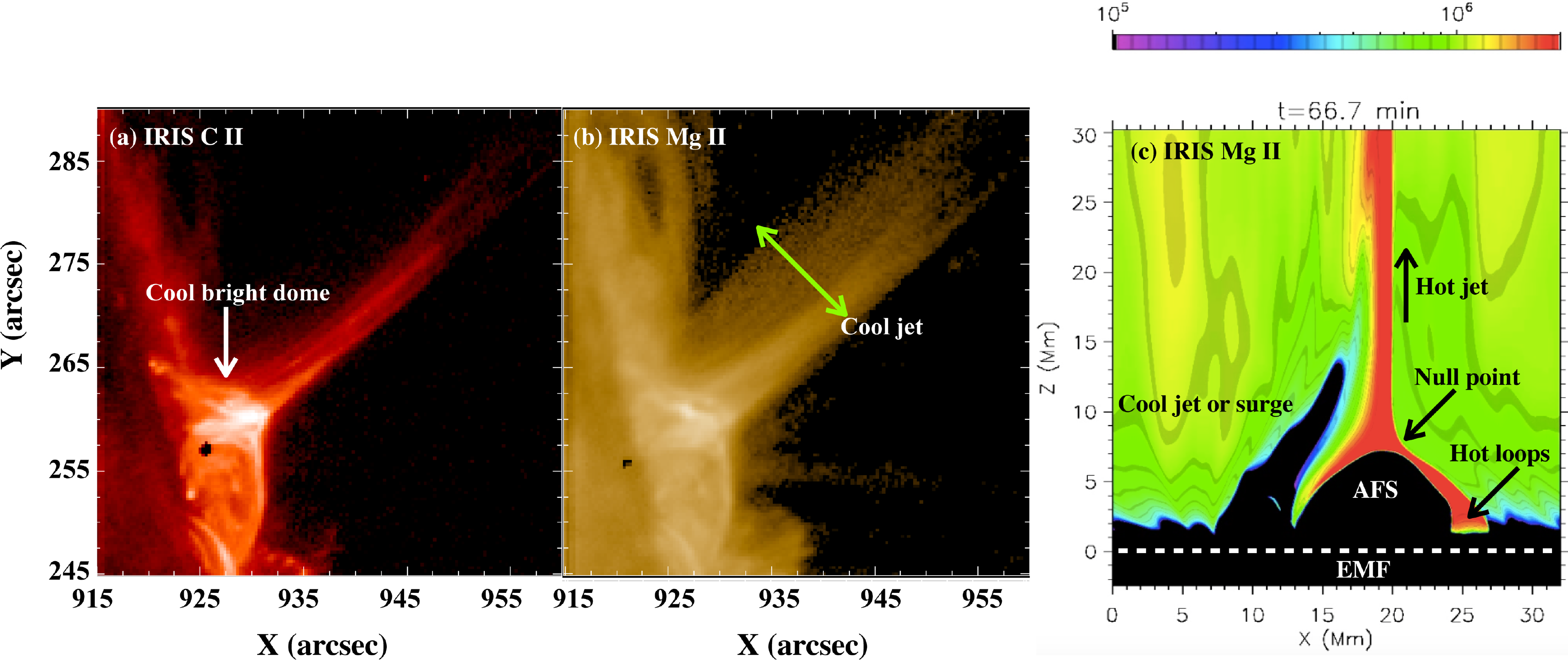}
\caption{IRIS jet observed in C II on April 4, 2017, showing a bright dome signature of emerging flux with a null point (left panel), in Mg II showing the large cool surge (middle panel)  (from \citet{Joshi2020}), (right panel) snapshot of the 3D  MHD simulation  of \citet{Nobrega2016}.
}
\label{Fig_nobrega}
\end{figure*}

\subsection{Case study of emerging flux at the limb viewed in  3D MHD  simulation}
Six recurrent jets  occurring in active region NOAA12644 on April 4, 2017. were observed in all the hot filters
of AIA as well as cool surges in IRIS slit-jaw high spatial and temporal resolution images \citep{Joshi2020}. 
The hot jets are collimated  ejections observed in the hot temperature  AIA filters (Figure \ref{Fig_nobrega}),  they have
high velocities (around 250 km/s) and are accompanied by cool surges and ejected kernels that both move at about 45 km/s.
This series of jets and surges provides a good case study for testing the 2D and 3D magnetohydrodynamic  (MHD) emerging
flux models (see the numerical simulations of \citet{Torok2009,Archontis2004,Moreno2008,Moreno2013}.  In their simulations they solved the
MHD equations in three dimensions to study the launching of
coronal jets following the emergence of magnetic flux.

The  jet observations at the limb inferr clearly a null point in the corona and a dome below in all the  AIA filters.
The double-chambered structure of the dome  corresponds to the regions with cold and hot loops that
are in the models below the current sheet that contains the reconnection site \citep{Nobrega2016}. The former model is based on the  
radiation-MHD Bifrost code \citep{Gudiksen2011}. In the 3D models, the jet is launched along open coronal field
lines that result from the reconnection of the emerged field with
the pre-existing ambient coronal field. Underneath the jet, two
vault structures are formed, one containing the emerging cool
plasma and the other a set of hot, closed coronal loops resulting
from the reconnection.
The cool surge with kernels is comparable with the cool 
ejection and plasmoids that naturally appears in the current  sheet in models \citep{Ni2021}.

The comparison
of the observations of the structures and time evolution of the
jet complex  observed at the limb with numerical experiments of the launching of jets
following flux emergence from below the photosphere shows significant  similarities, proving  that such a 3D experiment is valid to explain the AIA observations. Quantitatively the characteristics of the jets (speed and temperature) fit well with the values determined in the MHD simulations \citep{Nobrega2016}.
{ The  comparison of this  case study with  a model of emergence  suggests strongly  that this  jet  may have been  initiated  by flux emergence.}

{Another example of flux emergence was studied  by \citet{Yang2018}  showing a comprehensive force analysis of the cool and warm jets.   The   cool jet was mainly accelerated by the gradients of both thermal and magnetic pressures near the outer border of the mass-concentrated region, which is compressed by the emerging loop, while the hot  jet was accelerated mainly by the sling-shot effect (curvature tension of magnetic field) of reconnected magnetic field lines and heated directly  by  { resistive dissipation}.
}

\section{Untwisted jets and models}


\begin{figure*}[t!]
\centering
\includegraphics[width=0.8\textwidth]{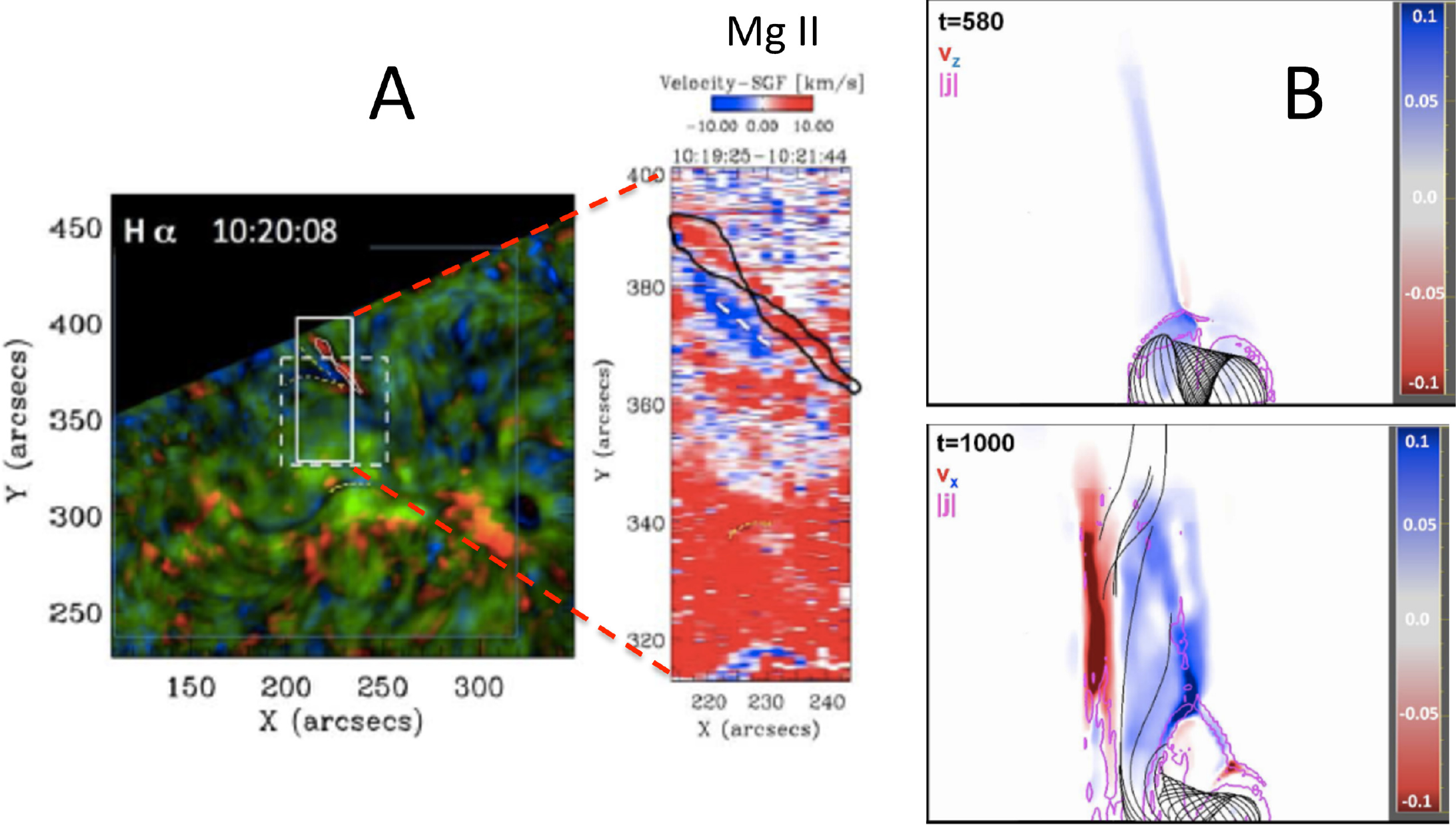}
\caption{Jet observation (A) in H$\alpha$  with the Meudon MSDP (left panel) in Mg II  with IRIS (middle panel) on March 30, 2017:
Dopplershift map (blue and red) combined with H$\alpha$ intensity (yellow and green) and the related Dopplershift raster map in  Mg II
(from \citet{Ruan2019}.
(B) Dopplermaps obtained for two times in the 3D MHD simulation of a twisted jet
showing that the opposite sides of the jet have opposite (red/blue) Dopplershifts (from \citet{Pariat2015}).
}
\label{Fig_pariat}
\end{figure*}

\subsection{Rotating structure}
Helical or rotating jets are frequently observed 
in AIA 304 \AA\  and in X-ray \citep{Nistico2009} { and in the multi channels of AIA \citep{Kumar2018}. }  Rotation rate, speed   have been estimated by following some fine structures  in the jet \citep{Chen2012,Hong2013}.
 Helical jets have been reconstructed in 3D by using STEREO spacecraft  \citep{Patsourakos2008}. 
 {  In the study of \citet{Schmieder2013} time-slice analysis  along  a jet  revealed a striped pattern of dark and bright strands propagating along its axis,  with
apparent damped oscillations across the jet   (Fig \ref{Fig_chandra} panel b). This was suggestive of a (un)twisting motion in the jet, possibly an Alfv\'en
wave.  Later, similar twisting was shown by other studies \citep{Panesar2016a}. { In some well-resolved, high-cadence observations, the untwisting itself and the helical structure were detected.
 Twist  has been observed in coronal-hole jets too (e.g., Kumar et al. 2018).} }

Helical  shape  and  twisting  in jets have  been  further  demonstrated by  advanced  spectroscopic methods.  
  Using IRIS spectra, Doppler-images revealed rotating jets showing blue and red shift on opposite sides of the jet axis  \citep{Jibben2004,Cheung2015,Ruan2019}.
Magnetic reconnection of twisted flux tubes with their less twisted surroundings can account for the production and rotating motion of the jets.
Cool jets or surges  show also such blue/red Dopplershifts parallel to the structure (Figure \ref{Fig_pariat} left panels) \citep{Tian2014,Ruan2019}.
{ Blue and redshift pattern is not always interpreted as a  possible rotation \citep{Schmieder1983,Tiwari2019},  However,  frequently  it is defined as a }
characteristics of twisting jet is relatively common and does not depend on the temperature or the coronal  environment. 
Twist has   been found in penumbra jets   \citep{Tiwari2018},  and
in active region jets 
\citep{Joshi2021_twist,Lu2019} using Si IV, C II  and Mg II lines observed by IRIS.  The interpretation of the small jets is based on   cartoons  showing  magnetic  reconnection in    mixed local magnetic field polarities. The spectra at the reconnection site show  bidirectional  flows either in the low chromosphere \citep{Joshi2021_twist}, either in the corona \citep{Ruan2019}.
or  at the top of an emergent mini-filament \citep{Tiwari2019}. 
{ Large Doppler  flows  can be   found  at the reconnection  e.g. around +/- 100-200 km/s \citep{Joshi2021_twist}.  Bi-directional flows have also been interpreted as  signature of  rotation in the jets themselves  \citep{Curdt2012,Pariat2016}.   However these Dopplershift  flows are measured  along the LoS  which generally is  nearly perpendicular to the  direction of the observed  jets. Therefore they correspond to reconnection-jets.}

\subsection{MHD models}
In \citet{Torok2009}, and \citet{Wyper2016}  helical jet consists
of untwisting upflows driven by the propagation of torsional
waves: these waves were induced by the sequential reconnection
of twisted closed field lines with the straight open
field.  The global picture is due to multiple sequential reconnections. In    numerical  models  of coronal jets generated in response to flux emergence,    
helical jets could be  driven by untwisting upflows e.g. \citep{Archontis2012}. 

In \citet{Pariat2009} the helical jet is released  by the  interchange reconnection between open and closed magnetic fields, which generates a series of impulsive nonlinear Alfvénic or kink waves. This kind of torsional  waves    propagate   with untwisting upflows  along  reconnection-formed  open  field  lines  and eject most of the twist (magnetic helicity) stored in the close domain (Figure \ref{Fig_pariat} right panels). 
In this model the close domain possess  a given  magnetic helicity  with 
close twisted field lines while in the emergence flux model  { the flux emerges already twisted or}  the twist is created by untwisting upflows.
{ In the \citet{Pariat2009} the twist is broadly distributed,  driving reconnection at  the breakout  current sheet without the formation of flux rope, while in \citet{Wyper2016}  the twist is concentrated along the PIL in a mini-filament    which forms an eruptive  flux rope.  Multiple reconnection sites are possible, below the FR and  at the breakout current sheet, as it was shown in  coronal hole jets \citep{Kumar2018}.}

\begin{figure*}[ht!]
\centering
\includegraphics[width=0.8\textwidth]{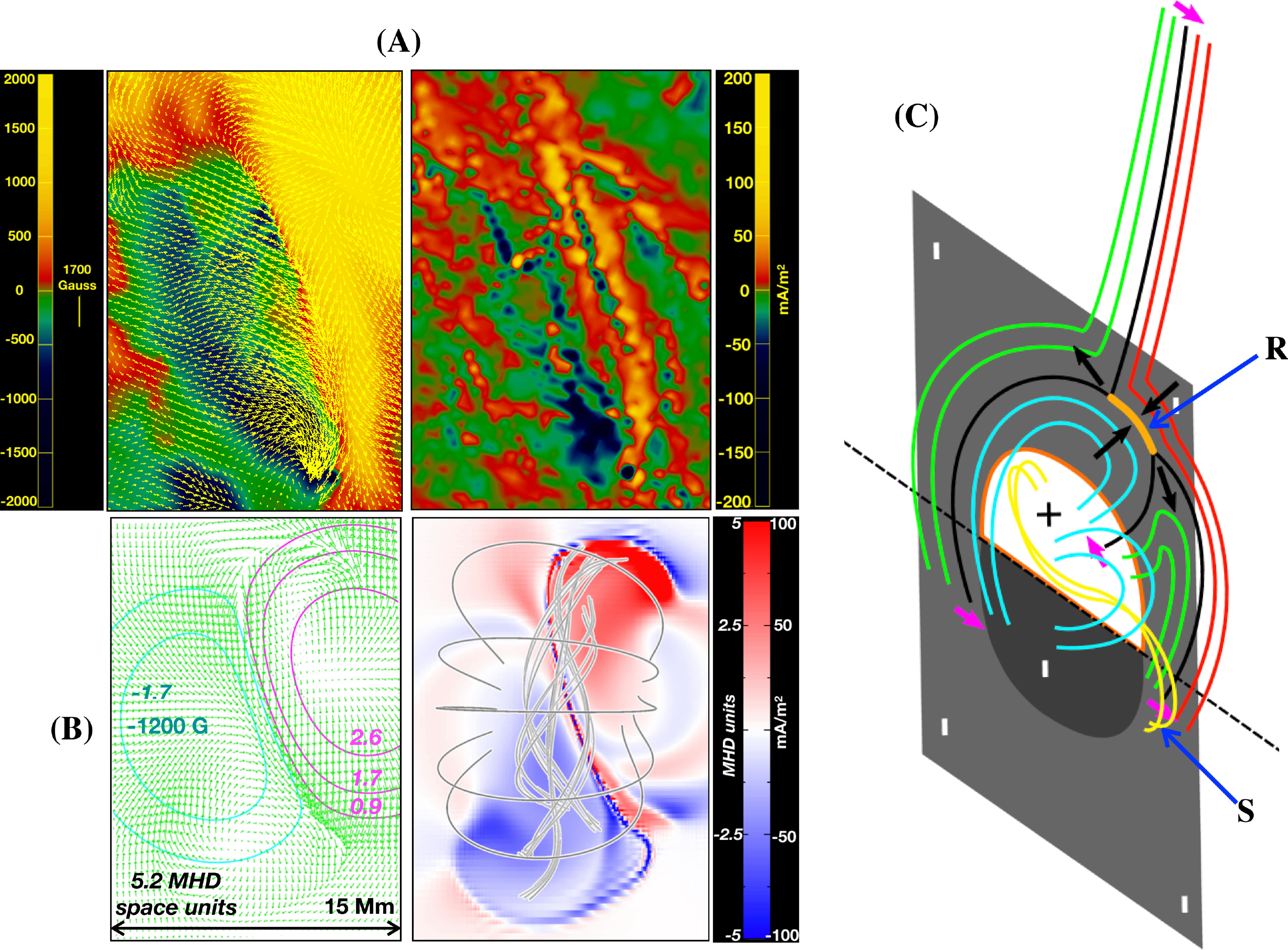}
\caption{Large flux rope detected in the HMI magnetic vector map computed with UNNOFIT code \citep{Bommier2016} (A) : magnetic field Bz overlaid by arrows of the horizontal magnetic field (the yellow (dark) blue areas show the positive (negative) magnetic field polarity) and  electric  current  density  map Jz; (B) comparison with the OHM MHD simulation of a flux rope. The vector pattern of observations and model looks the same, as they are strongly nearly parallel to the PIL and converging together in the bottom part to the site of reconnection S.  (from \citet{Joshi2020FR}); (C) schematic view   of magnetic field lines in the jet bald patch   MHD simulation; in yellow is drawn  the flux rope with a  sigmoid shape (adapted from \citet{Wyper2019}). { The reconnection is  at R  in the paper of \citet{Wyper2019} but , a reconnection at S will  be a better  fit to  the  \citet{Joshi2020FR}  observations.}
}
\label{Fig_FR}
\end{figure*}

\subsection{Transfer of twist}

A new interesting case study was provided by a jet observed on March 22 2019  with SDO and IRIS studied by two groups using different tools \citep{Joshi2020FR,Yang2020}. The  active region  was formed by a series of emerging flux, which evolved very rapidly and produced many micro flares. At the time of the jet new emerging flux  squeezed next to    formerly  emerged flux. The leading negative polarity of the bipole  slipped along the following negative polarity of the older bipole \citep{Joshi2020FR}.  This scenario  favours  reconnection \citep{Syntelis2015}.
At the time  of the jet a part of the two  bipoles close to each other  fragmented and the jet occurred just at this smaller, newly created  bipole.
Therefore  the jet occurred between two arch filament systems which reconnected in a bald patch region. The bald patch region is transformed dynamically into a  null point  within a  current sheet and a twisted  jet is expelled. This model of bald patch  eruption has been studied by \cite{Wyper2019} (Figure \ref{Fig_FR} right panel).
The question is,  where do  the energy that  powers the jet and the twist of the jet. come from?
The vector magnetograms unveil the existence of a large flux rope with a sigmoid shape (Figure \ref{Fig_FR} left panels). Electric currents are detected in the hooks. The flux rope carries  the energy  release  during the reconnection, and its  twist is transferred during the reconnection.  \citet{Joshi2020FR}  compared these observations to the results of a numerical simulation of flux rope \citet{Aulanier2010,Zuccarello2015}.
 The observed vector  magnetic-field vector  pattern and the values of the vertical  electric current density are in good agreement
 with synthetic  vertical electric current density and vector B data from the MHD flux rope (FR) model, which reveals the FR location.
The Mg II spectra observed  at the base of the jet show a bi-directional extension with flows reaching 300 km/s.
The spectra along the slit show a slow decrease of the velocity along  the slit  as it crosses the main section of the jet, proving  that the jet is rotating   \citep{Joshi2021_twist}.

 The second group explains  this event  differently   \citep{Yang2020}. They performed a non linear force free extrapolation  and identified  a null point, a fan and a long spine. 
 This configuration differs from that inferred by \citet{Joshi2020FR}.
  They observed a small H$\alpha$ filament  with  the  high resolution   NVST telescope in the middle of the small bipole  and suggested that this filament (FR) has a role in  triggering the jet. They identified a second filament which does not  correspond to  the large  FR found by  the other group.  As it was mentioned earlier it is difficult to distinguish a filament from an arch filament system. This could be the case of this second filament which has  in fact no role in their interpretation.  They proposed a break-out model which might
remove the overlaying arcades, leaving the small FR to erupt and
turn into a blowout jet as in the scenario of \citet{Sterling2015}.
This jet is  explained by  the  breakout model  for jets with mini-filament  \citep{Wyper2018}.
The conclusion of these two studies is that it is again difficult to understand the real  driver of  the jets and surges.  {  In the  \citet{Joshi2020FR}  the twist was transferred from  a distant FR  experimenting
 fragmentation  while  in \citet{Yang2020} the twist came directly from a mini-filament  observed at the limit of the telescope resolution. Both interpretations  are  interesting.  }
 A data driven study could help to understand the evolution of the active region  magnetic topology leading to this jet and other jets.  
On the theoretical point of view data driven simulations start to be very promising closer to the observations they may unveil the secret of  jets \citep{Guo2021}.

\section{Conclusion and Perspective}
 Solar jets have been observed for  more  than 50 years in multi wavelengths with steadily increasing  spatial and temporal resolution instruments from    ground based and space telescopes. 
 Jet characteristics (length, speed, width) span  large domains of ranges. They are observed all over the solar disk, in active region, in coronal holes,  mainly at the edge of close structures neighbouring  open structures or large loops.
 Several recent partial or   complete reviews exist on this topics \citep{Innes2016,Raouafi2016,Hinode2019,Pariat2016,Shen2021,Schmieder2022}.
 
In this paper we  approach  this subject with a critical view than differs from that of  the previous reviews. Following  a chronological  order we  constantly present observations (SDO and IRIS) and the associated  theoretical models, either with cartoons 
or 2D and 3D  MHD simulations as they  developed progressively.  Substantial progress  has  been achieved concerning the analysis of the magnetic topology of jets.
The results can be summarised as follows:
\begin{enumerate}
    \item 
Magnetic reconnection  triggers  surges and  jets and  could  occur in electric current layers   associated with  null point, bald patch, separatrices, QSLs.
\item { Convection is the main  force which initiates  photospheric motions leading to shear, flux cancellation,   flux emergence and consequently magnetic reconnection.}
\item
Electric current layers form between two different magnetic systems e.g.   emerging  magnetic flux and overlaying magnetic field. An intrusion of opposite polarity is  in generally detected in the magnetograms at the base of the jet. { However intrusions of opposite polarity are difficult to detect in quiet-Sun and coronal-hole magnetograms, because the LoS fields are weak there and are close to the HMI lower limit. DKIST should be more sensitive and might be able to demonstrate more conclusively whether flux cancellation is occurring. } 
\item
Kinetic energy  of  reconnection jets comes from the dissipation of the magnetic field in the current sheet. 
{ The dissipation favours the changes of  geometry of  field lines generating strong curvatures in the field lines.}
 Therefore a tension force operates, and  the system grows. The jets have Alfv\'enic speed because the majority of  the dissipated magnetic energy  (and even 100\% in the Sweet Parker type models) is converted into kinetic energy.
\item
Twisted jets are frequently observed.  { Twist should be  already present in the closed region, either by kinking, flux rope formation, emergence of preexisting twisted flux, or post-emergence rotation. The twist has to be  transferred to the jet-hosting field lines through reconnection.  }
\end{enumerate}
Kinetic energy from the untwisting jet comes from the reconnection between a twisted force free field (fff) loop with an un-twisted fff loop. The reconnected loop is twisted on one side and not on the other, it generates a non-fff at the interface, and therefore the twist will be distributed by means of J $\times$ B along the field, a force which therefore also pushes the plasma in the  direction of twisted field lines around  and  along the field lines. The speeds here depend on the magnitude of J $\times$ B. so it may be different from Alfvenic speed.\\
Many questions about jets still stay open and need to be clarified. They are the seeds of many important questions relative to coronal heating \citep{Berghmans2021,Panesar2021},  sources of the solar wind \citep{Fargette2021},  acceleration of particles \citep{Pick2006,Wang2006,Joshi2021b},  narrow coronal mass ejections \citep{Shen2012,Kumar2021,Panesar2016b}.
Coordinated observations with new spacecraft (Parker Solar  Probe  - \citet{Fox2016} and Solar Orbiter- \citet{Muller2020}) and high ground based instruments e.g. DKIST, EST will favour a breakthrough in  our knowledge of solar jets and their related phenomena.

\section*{Acknowledgments}
IRIS is a NASA small explorer mission developed and operated by LMSAL with mission
operations executed at NASA Ames Research center and major contributions to downlink
communications funded by the Norwegian Space Center (NSC, Norway) through an ESA
PRODEX contract. We thank the SDO/AIA, SDO/HMI, and IRIS science teams for granting free access to the data. 
The author acknowledge the three referees who help her to improve the paper substantially. 
\section*{Conflict of Interest Statement}

The author declares that the research was conducted in the absence of any commercial or financial relationships that could be construed as a potential conflict of interest.

\section*{Author Contributions}
BS declares to be the sole author. BS  thanks Reetika Joshi for achieving a few Figures.
BS thanks Drs K.Shibata, S.Tiwari, A.Sterling  and   Guillaume Aulanier for fruitful discussions on the theory of jets.

\bibliographystyle{frontiersinSCNS_ENG_HUMS} 




\end{document}